\documentclass[showpacs]{revtex4}
\usepackage{indentfirst}
\begin{document}
\title{\bf Heavy dibaryons}
\author{S.M. Gerasyuta}
\email{gerasyuta@SG6488.spb.edu}
\author{E.E. Matskevich}
\email{matskev@pobox.spbu.ru}
\affiliation{Department of Theoretical Physics, St. Petersburg State University, 198904,
St. Petersburg, Russia}
\affiliation{Department of Physics, LTA, 194021, St. Petersburg, Russia}
\begin{abstract}
The relativistic six-quark equations are found in the framework of the dispersion
relation technique. The approximate solutions of these equations using the method
based on the extraction of leading singularities of the heavy hexaquark
amplitudes are obtained. The relativistic six-quark amplitudes of dibaryons
including the light quarks $u$, $d$ and heavy quarks $c$, $b$ are calculated.
The poles of these amplitudes determine the masses of charmed and bottom dibaryons
with the isospins $\frac{1}{2}$, $\frac{3}{2}$, $\frac{5}{2}$.
\end{abstract}
\pacs{11.55.Fv, 12.39.Ki, 12.39.Mk, 12.40.Yx.}
\maketitle
\section{Introduction.}
Hadron spectroscopy has always played an important role in the revealing
mechanisms underlying the dynamic of strong interactions.

The heavy hadron containing a single heavy quark is particularly
interesting. The light degrees of freedom (quarks and gluons) circle
around the nearby static heavy quark. Such a system behaves as the QCD
analog of familar hydrogen bound by the electromagnetic interaction.

The heavy quark expansion provides a systematic tool for heavy hadrons.
When the heavy quark mass $m_Q \to \infty$, the angular momentum of the
light degree of freedom is a good quantum number. Therefore, heavy hadrons
form doublets. For example, $\Omega_b$ and $\Omega^*_b$ will be degenerate
in the heavy quark limit. Their mass splitting is caused by the chromomagnetic
interaction at the order $O(1/m_Q)$, which can be taken into account
systematically in the framework of heavy quark effective field theory (HQET)
\cite{1, 2, 3}.

In 1977, Jaffe \cite{4} studied the color-magnetic interaction of the
one-gluon-exchange potential in the multiquark system and found that the
most attractive channel is the flavor singlet with quark content
$u^2d^2s^2$. The same symmetry analysis of the chiral boson exchange
potential leads to the similar result \cite{5}.

The $H$-particle, $N\Omega$-state and di-$\Omega$ may be strong interaction
stable. Up to now, these three interesting candidates of dibaryons are still
not found or confirmed by experiments. It seems that one should go beyond
these candidates and should search the possible candidates in a wider region,
expecially the systems with heavy quarks, in terms of a more reliable model.

There were a number of theoretical predictions by using various models:
the quark cluster model \cite{6, 7}, the quark-delocation model \cite{8, 9}, the
chiral $SU(3)$ quark model \cite{10}, the flavor $SU(3)$ skyrmion model \cite{11}.
Lomon predicted a deuteronlike dibaryon resonance using R-matrix theory
\cite{12}. By employing the chiral $SU(3)$ quark model Zhang and Yu studied
$\Omega\Omega$ and $\Sigma\Omega$ states \cite{13, 14}.

In a series of papers \cite{15, 16, 17, 18, 19} a method has been developed which is
convenient for analyzing relativistic three-hadron systems. The physics of
the three-hadron system can be described by means of a pair interaction
between the particles. There are three isobar channels, each of which
consists of a two-particle isobar and the third particle. The presence
of the isobar representation together with the condition of unitarity in
the pair energies and of analyticity leads to a system of integral equations
in a single variable. Their solution makes it possible to describe the
interaction of the produced particles in three-hadron systems.

In Ref. \cite{20} a representation of the Faddeev equation in the form of a
dispersion relation in the pair energy in the two interacting particles
was used. This was found to be convenient in order to obtain an approximate
solution of the Faddeev equation by a method based on extraction of the
leading singularities of the amplitude. With a rather crude approximation
of the low-energy $NN$ interaction a relatively good description of the
form factor of tritium (helium-3) at low $q^2$ was obtained.

In our papers \cite{21, 22, 23} relativistic generalization of the three-body
Faddeev equations was obtained in the form of dispersion relations in the
pair energy of two interacting quarks. The mass spectrum of $S$-wave
baryons including $u$, $d$, $s$ quarks was calculated by a method based on
isolating the leading singularities in the amplitude. We searched for the
approximate solution of integral three-quark equations by taking into
account two-particle and triangle singularities, all the weaker ones being
neglected. If we considered such approximation, which corresponds to
taking into account two-body and triangle singularities, and defined all
the smooth functions of the subenergy variables (as compared with the
singular part of the amplitude) in the middle point of the physical region
of Dalitz-plot, then the problem was reduced to the one of solving a system
of simple algebraic equations.

In the present paper the relativistic six-quark equations are found in
the framework of coupled-channel formalism. We use only planar diagrams; the other
diagrams due to the rules of $1/N_c$ expansion \cite{24, 25, 26} are neglected.

The six-quark amplitudes of dibaryons are calculated. The poles of these amplitudes
determine the masses of dibaryons. We calculated the contribution of six-quark
subamplitudes to the hexaquark amplitudes. In Sec. II, we briefly discuss
the relativistic Faddeev approach. The relativistic three-quark equations
are constructed in the form of the dispersion relation over the two-body
subenergy. The approximate solution of these equations using the method
based on the extraction of leading singularities of the amplitude are
obtained. We calculated the mass spectrum of $S$-wave bottom baryons with
$J^P=\frac{1}{2}^+$, $\frac{3}{2}^+$ (Tables \ref{tab1}, \ref{tab2}).
In Sec. III, the six-quark amplitudes of hexaquarks are constructed. The
dynamical mixing between the subamplitudes of dibaryons are considered.
The relativistic six-quark equations are constructed in the form of the
dispersion relation over the two-body subenergy. The approximate solutions
of these equations using the method based on the extraction of leading
singularities of the amplitude are obtained. Sec. IV is devoted to the
calculation results for the dibaryon mass spectra and the contributions of
subamplitudes to the hexaquark amplitude (Tables \ref{tab4}, \ref{tab5},
\ref{tab6}, \ref{tab7}). In conclusion, the status of the considered
model is discussed.

\section{Brief introduction of relativistic Faddeev equations.}

In our papers, \cite{21, 22, 23, 27, 28} relativistic generalization of the
three-body Faddeev equations was obtained in the form of dispersion relations
in the pair energy of two interacting particles. The mass spectra of $S$-wave
baryons including $u$, $d$, $s$, $c$ quarks were calculated by a method
based on isolating of the leading singularities in the amplitude.

We searched for the approximate solution of integral three-quark equations
by taking into account two-particle and triangle singularities, the
weaker ones being neglected. If we considered such an approximation, which
corresponds to taking into account two-body and triangle singularities, and
defined all the smooth functions at the middle point of the physical region
of Dalitz-plot, then the problem was reduced to the one of solving a system
of simple algebraic equations.

In the paper \cite{27} the relativistic three-particle amplitudes in the
coupled-channels formalism are considered. We take into account the
$u$, $d$, $s$, $c$, $b$ quarks and construct the flavor-spin functions
for the $35$ baryons with the spin-parity $J^p=\frac{1}{2} ^{+}$ and
$J^p=\frac{3}{2} ^{+}$:

\vskip2ex

$$
\begin{tabular}{llp{30ex}ll}
 & $J^p=\frac{1}{2} ^{+}$ & & & $J^p=\frac{3}{2} ^{+}$ \\
 & & & & \\
$\Sigma_b$ & $uub, udb, ddb$ & & $\Sigma_b$ & $uub, udb, ddb$ \\
$\Lambda_b$ & $udb$ & & $\Xi_{sb}$ & $usb, dsb$ \\
$\Xi_{sb}^A$ & $usb, dsb$ & & $\Omega_{ssb}$ & $ssb$ \\
$\Xi_{sb}^S$ & $usb, dsb$ & & $\Xi_{cb}$ & $ucb, dcb$ \\
$\Omega_{ssb}$ & $ssb$ & & $\Omega_{scb}$ & $scb$ \\
$\Xi_{cb}^A$ & $ucb, dcb$ & & $\Omega_{ccb}$ & $ccb$ \\
$\Xi_{cb}^S$ & $ucb, dcb$ & & $\Xi_{bb}$ & $ubb, dbb$ \\
$\Lambda_{scb}^A$ & $scb$ & & $\Omega_{sbb}$ & $sbb$ \\
$\Lambda_{scb}^S$ & $scb$ & & $\Omega_{cbb}$ & $cbb$ \\
$\Omega_{ccb}$ & $ccb$ & & $\Omega_{bbb}$ & $bbb$ \\
$\Xi_{bb}$ & $ubb, dbb$ & & & \\
$\Omega_{sbb}$ & $sbb$ & & & \\
$\Omega_{cbb}$ & $cbb$ & & & \\
\end{tabular}
$$

\vskip2ex

In the paper \cite{28}, the relativistic equations were obtained and the mass
spectrum of $S$-wave charmed baryons was calculated.

In the present paper, we will be able to use the similar method. In this
case, we consider $35$ baryons with the spin-parity
$J^p=\frac{1}{2} ^{+}$ and $J^p=\frac{3}{2} ^{+}$, which include one, two
and three bottom quarks. We have considered the $23$ baryons with different
masses (Tables \ref{tab1}, \ref{tab2}).

We calculate the masses of the bottom baryons in a relativistic approach
using the dispersion relation technique. The relativistic three-quark
integral equations are constructed in the form of the dispersion relations
over the two-body subenergy.

We use the graphical equations for the functions $A_J(s,s_{ik})$.
In order to represent the amplitude $A_J(s,s_{ik})$ in the form
of dispersion relations, it is necessary to define the amplitudes of
quark-quark interaction $a_J(s_{ik})$. The pair amplitudes
$qq\rightarrow qq$ are calculated in the framework of the dispersion
$N/D$ method with the input four-fermion interaction with quantum numbers
of the gluon. We use results of our relativistic quark model \cite{29}
and write down the pair quark amplitudes in the following form:

\begin{equation}
a_J(s_{ik})=\frac{G^2_J(s_{ik})}
{1-B_J(s_{ik})} \, ,\end{equation}

\begin{equation}
\label{2}
B_J(s_{ik})=\int\limits_{(m_i+m_k)^2}^{\Lambda_J (i,k)}\hskip2mm
\frac{ds'_{ik}}{\pi}\frac{\rho_J(s'_{ik})G^2_J(s'_{ik})}
{s'_{ik}-s_{ik}} \, ,\end{equation}

\begin{eqnarray}
\rho_J (s_{ik})&=&\frac{(m_i+m_k)^2}{4\pi}
\left(\alpha_J\frac{s_{ik}}{(m_i+m_k)^2}
+\beta_J +\frac{\delta_J}{s_{ik}} \right)
\nonumber\\
&&\nonumber\\
&\times &\frac{\sqrt{(s_{ik}-(m_i+m_k)^2)(s_{ik}-(m_i-m_k)^2)}}
{s_{ik}} \, .
\end{eqnarray}

Here $G_J$ is the vertex function of a diquark, which can be expressed in
terms of the $N$-function of the bootstrap $N/D$ method as $G_J=\sqrt{N_J}$,
$B_J(s_{ik})$ is the Chew-Mandelstam function \cite{30}, and $\rho_J (s_{ik})$
is the phase space for a diquark. $s_{ik}$ is the pair energy squared of
diquark, the index $J^p$ determines the spin-parity of diquark.
The coefficients of Chew-Mandelstam function $\alpha_J$, $\beta_J$ and
$\delta_J$ in Table \ref{tab3} are given. $\Lambda_J(i,k)$ is the pair energy
cutoff. In the case under discussion the interacting pairs of quarks do not
form bound states. Therefore, the integration in the dispersion integral (\ref{2})
is carried out from $(m_i+m_k)^2$ to $\Lambda_J(i,k)$ ($i,k=1,2,3$). Including
all possible rescatterings of each pair of quarks and grouping the terms
according to the final states of the particles, we obtained the coupled
systems of integral equations. For instance, for the
$\Sigma^+_b$ with $J^p=\frac{1}{2} ^{+}$ the wave function is
$\varphi_{\Sigma^+_b}=\sqrt{\frac{2}{3}}\{u\uparrow d\uparrow b\downarrow\}-
\sqrt{\frac{1}{6}}\{u\uparrow d\downarrow b\uparrow\}-
\sqrt{\frac{1}{6}}\{u\downarrow d\uparrow b\uparrow\}$. Then the coupled
system of equations has the following form:

\begin{eqnarray}
\label{4}
\left\{
\begin{array}{l}
A_1(s,s_{12})=\lambda\, b_1(s_{12})L_1(s_{12})+
K_1(s_{12})\left[\frac{1}{4}A_{1^b}(s,s_{13})+
\frac{3}{4}A_{0^b}(s,s_{13})+\right.\\
\\
\hskip10ex \left.
+\frac{1}{4}A_{1^b}(s,s_{23})+\frac{3}{4}A_{0^b}(s,s_{23})
\right]\\
\\
A_{1^b}(s,s_{13})=\lambda\, b_{1^b}(s_{13})L_{1^b}(s_{13})+
K_{1^b}(s_{13})\left[\frac{1}{2}A_1(s,s_{12})-
\frac{1}{4}A_{1^b}(s,s_{12})+\right.\\
\\
\hskip10ex \left.
+\frac{3}{4}A_{0^b}(s,s_{12})+\frac{1}{2}A_1(s,s_{23})-
\frac{1}{4}A_{1^b}(s,s_{23})+\frac{3}{4}A_{0^b}(s,s_{23})
\right] \\
\\
A_{0^b}(s,s_{23})=\lambda\, b_{0^b}(s_{23})L_{0^b}(s_{23})+
K_{0^b}(s_{23})\left[\frac{1}{2}A_1(s,s_{12})+
\frac{1}{4}A_{1^b}(s,s_{12})+\right.\\
\\
\hskip10ex \left.
+\frac{1}{4}A_{0^b}(s,s_{12})+\frac{1}{2}A_1(s,s_{13})+
\frac{1}{4}A_{1^b}(s,s_{13})+\frac{1}{4}A_{0^b}(s,s_{13})
\right] \, .\\
\end{array} \right.
\end{eqnarray}

\noindent
Here the function $L_J(s_{ik})$ has the form

\begin{equation}
L_J(s_{ik})=\frac{G_J(s_{ik})}{1-B_J(s_{ik})}
 \, .\end{equation}

\noindent
The integral operator $K_J (s_{ik})$ is

\begin{equation}
K_J (s_{ik})=L_J(s_{ik})\, \int\limits_{(m_i+m_k)^2}^{\Lambda_J(ik)}
\hskip2mm\frac{ds'_{ik}}{\pi}\frac{\rho_J(s'_{ik})G_J(s'_{ik})}
{s'_{ik}-s_{ik}}\, \int\limits_{-1}^{1}\frac{dz}{2}
 \, .\end{equation}

\noindent
The function $b_J(s_{ik})$ is the truncated function of Chew-Mandelstam:

\begin{equation}
b_J(s_{ik})=\int\limits_{(m_i+m_k)^2}^{\infty}\hskip2mm
\frac{ds'_{ik}}{\pi}\frac{\rho_J(s'_{ik})G_J(s'_{ik})}
{s'_{ik}-s_{ik}}
 \, ,\end{equation}

\noindent
$z$ is the cosine of the angle between the relative momentum of particles
$i$ and $k$ in the intermediate state and the momentum of
particle $j$ in the final state, taken in the c.m. of the particles
$i$ and $k$. Let some current produces three quarks
with the vertex constant $\lambda$. This constant do not affect to the
spectra mass of bottom baryons. By analogy with the $\Sigma^+_b$ state, we obtain
the rescattering amplitudes of the three various quarks for the all
bottom states.

Let us extract two-particle singularities in $A_J(s,s_{ik})$:

\begin{equation}
A_J(s,s_{ik})=\frac{\alpha_J(s,s_{ik})b_J(s_{ik})G_J(s_{ik})}
{1-B_J(s_{ik})}
 \, ,\end{equation}

\noindent
$\alpha_J(s,s_{ik})$ is the reduced amplitude. Accordingly to this,
all integral equations can be rewritten using the reduced amplitudes.
The function $\alpha_J(s,s_{ik})$ is the smooth function of $s_{ik}$
as compared with the singular part of the amplitude. We do not extract
the three-body singularities, because they are weaker than the
two-particle singularities. For instance, one considers the first equation
of system for the $\Sigma^+_b$ with $J^p=\frac{1}{2}^+$:

\begin{eqnarray}
\alpha_1 (s,s_{12})&=&\lambda+\frac{1}{b_1(s_{12})}
\, \int\limits_{(m_1+m_2)^2}^{\Lambda_1(1,2)}\hskip2mm
\frac{ds'_{12}}{\pi}\,\frac{\rho_1(s'_{12})G_1(s'_{12})}
{s'_{12}-s_{12}}\nonumber\\
&&\nonumber\\
&\times & \int\limits_{-1}^{1}\frac{dz}{2}\,
\left(
\frac{G_{1^b}(s'_{13})b_{1^b}(s'_{13})}{1-B_{1^b}(s'_{13})}
\,\frac{1}{2}\,\alpha_{1^b}(s,s'_{13})+
\frac{G_{0^b}(s'_{13})b_{0^b}(s'_{13})}{1-B_{0^b}(s'_{13})}
\,\frac{3}{2}\,\alpha_{0^b}(s,s'_{13})
\right).
\end{eqnarray}

The connection between $s'_{12}$ and $s'_{13}$ is \cite{21}:

\begin{eqnarray}
\label{10}
s'_{13}&=&m_1^2+m_3^2-\frac{\left(s'_{12}+m_3^2-s\right)
\left(s'_{12}+m_1^2-m_2^2\right)}{2s'_{12}}\nonumber\\
&\pm & \frac{z}{2s'_{12}}
\sqrt{\left(s'_{12}-(m_1+m_2)^2\right)
\left(s'_{12}-(m_1-m_2)^2\right)
\left(s'_{12}-(\sqrt{s}+m_3)^2\right)
\left(s'_{12}-(\sqrt{s}-m_3)^2\right)}\, .
\end{eqnarray}

The formula for $s'_{23}$ is similar to (\ref{10}) with replaced by $z \to -z$.
Thus $A_{1^b}(s,s'_{13})+A_{1^b}(s,s'_{23})$ must be replaced by
$2A_{1^b}(s,s'_{13})$. $\Lambda_J(i,k)$ is the cutoff at the large
value of $s_{ik}$, which determines the contribution from small distances.

The construction of the approximate solution of coupled system-equations
is based on the extraction of the leading singularities which are close to
the region $s_{ik}=(m_i+m_k)^2$ \cite{31}.

We consider the approximation, which corresponds to the single interaction
of the all three particles (two-particle and triangle singularities) and
neglecting all the weaker ones.

The functions $\alpha_J(s,s_{ik})$ are the smooth functions of $s_{ik}$
as compared with the singular part of the amplitude, hence it can be
expanded in a series at the singulary point and only the first term of
this series should be employed further. As $s_0$ it is convenient to
take the middle point of physical region of the Dalitz-plot in which $z=0$.
In this case, we get from (\ref{10})
$s_{ik}=s_0=\frac{s+m_1^2+m_2^2+m_3^2}{m_{12}^2+m_{13}^2+m_{23}^2}$,
where $m_{ik}=\frac{m_i+m_k}{2}$. We define functions $\alpha_J(s,s_{ik})$
and $b_J(s_{ik})$ at the point $s_0$. Such a choice of point $s_0$ allows us
to replace integral equations (\ref{4}) by the algebraic couple equations for the
state $\Sigma^+_b$:

\begin{eqnarray}
\left\{
\begin{array}{l}
\alpha_1(s,s_0)=\lambda+\frac{1}{2}\,\alpha_{1^b}(s,s_0)
\, I_{1 1^b}(s,s_0)\,\frac{b_{1^b}(s_0)}{b_1(s_0)}
+\frac{3}{2}\,\alpha_{0^b}(s,s_0)\, I_{1 0^b}(s,s_0)
\,\frac{b_{0^b}(s_0)}{b_1(s_0)}\\
\\
\alpha_{1^b}(s,s_0)=\lambda
+\alpha_1(s,s_0)\, I_{1^b 1}(s,s_0)\,\frac{b_1(s_0)}{b_{1^b}(s_0)}\\
\\
\hskip10ex
-\frac{1}{2}\,\alpha_{1^b}(s,s_0)\, I_{1^b 1^b}(s,s_0)
+\frac{3}{2}\,\alpha_{0^b}(s,s_0)\, I_{1^b 0^b}(s,s_0)
\,\frac{b_{0^b}(s_0)}{b_{1^b}(s_0)}\\
\\
\alpha_{0^b}(s,s_0)=\lambda+\alpha_1(s,s_0)\, I_{0^b 1}(s,s_0)
\,\frac{b_1(s_0)}{b_{0^b}(s_0)}\\
\\
\hskip10ex
+\frac{1}{2}\,\alpha_{1^b}(s,s_0)\, I_{0^b 1^b}(s,s_0)
\,\frac{b_{1^b}(s_0)}{b_{0^b}(s_0)}
+\frac{1}{2}\,\alpha_{0^b}(s,s_0)\, I_{0^b 1^b}(s,s_0)
\, .\\
\end{array} \right.
\end{eqnarray}

The function $I_{J_1 J_2}(s,s_0)$ takes into account singularity
which corresponds to the simultaneous vanishing of all propagators in the
triangle diagram.

\begin{equation}
I_{J_1 J_2}(s,s_0)=\int\limits_{(m_i+m_k)^2}^{\Lambda_{J_1}}\hskip2mm
\frac{ds'_{ik}}{\pi}\frac{\rho_{J_1}(s'_{ik})G^2_{J_1}(s'_{ik})}
{s'_{ik}-s_{ik}}\, \int\limits_{-1}^{1}\frac{dz}{2}\,
\frac{1}{1-B_{J_2}(s_{ij})} \, .
\end{equation}

The $G_J(s_{ik})$ functions have the smooth dependence from energy
$s_{ik}$ \cite{27}, therefore we suggest them as constants. The parameters of
model: $g_J$ vertex constant and $\lambda_J$ cutoff parameter are chosen
dimensionless;

\begin{equation}
\label{13}
g_J=\frac{m_i+m_k}{2\pi}G_J , \qquad \lambda_J=\frac{4\Lambda_J}
{(m_i+m_k)^2} \, .
\end{equation}

Here $m_i$ and $m_k$ are quark masses in the intermediate state of the quark
loop. We calculate the coupled system of equations and can determine the
mass values of the $\Sigma^+_b$ state. We calculate a pole in $s$
which corresponds to the bound state of the three quarks.

By analogy with $\Sigma^+_b$-hyperon we obtain the systems of equations for
the reduced amplitudes of all bottom baryons.

The solutions of the coupled system of equations are considered as:

\begin{equation}
\alpha_J=\frac{F_J(s,\lambda_J)}{D(s)}\, ,
\end{equation}

\noindent
where the zeros of the $D(s)$ determinate the masses of bound
states of baryons. $F_J(s,\lambda_J)$ are the functions of $s$ and
$\lambda_J$. The functions $F_J(s,\lambda_J)$ determine the contributions
of subamplitudes to the baryon amplitude.

In quark models, which describe rather well the masses and static properties
of hadrons, the masses of the quarks usually have the similar values for
the spectra of light and heavy baryons. However, this is achieved at the
expense of some difference in the characteristics of the confinement
potential. It should be borne in mind that for a fixed hadron mass the
masses of the dressed quarks which enter into the composition of the hadron
will become smaller when the slope of the confinement potential increases
or its radius decreases. Therefore, conversely, we can change the masses
of the dressed quarks when going from the spectrum of light baryons to the
heavy baryons, while keeping the characteristics of the confinement
potential unchanged. We can effectively take into account the contribution
of the confinement potential in obtaining the spectrum of $S$-wave heavy
baryons.

In the case of $b$ quark we have used two new parameters: the cutoff of the $bb$
diquark  $\lambda_b=5.4$ and the coupling constant $g_b=1.03$.
These values have been determined by the $b$-baryon masses:
$M_{\Sigma_b \frac{1}{2}^+}=5.808\, GeV$ and
$M_{\Sigma_b \frac{3}{2}^+}=5.829\, GeV$. In order to fix $m_b=4.840\, GeV$
we use the $b$-baryon mass $5.829\, GeV$. We represent the masses of all
$S$-wave bottom baryons in the Tables \ref{tab1}, \ref{tab2}.
The calculated mass value $M_{\Lambda_b \frac{1}{2}^+}=5.624\, GeV$ is equal
to the experimental data \cite{32}, the mass value
$M_{\Xi^A_{sb} \frac{1}{2}^+}=5.761\, GeV$ is close to the experimental
one. We neglect with the mass distinction of $u$ and $d$ quarks.
The estimation of the theoretical error on the bottom baryon masses is
$2-5\, MeV$. This result was obtained by the choice of model parameters.

In our model the spin-averaged mass of the states $\Xi'_b$
and $\Xi^*_b$ is predicted to lie around to $250\, MeV$ above $M_{\Xi_b}$.
The relativistic corrections are particularly important for the splitting
between $\Omega^+_b$ and $\Omega_b$ baryons.

In the context of $\Xi'_b$ and $\Xi^*_b$ masses, it is worth mentioning
two relations among bottom baryons which incorporate the effects of
$SU(3)_f$ breaking:

\begin{equation}
\left( M\left( \Sigma^*_b \right)-M\left( \Sigma_b \right)\right)+
\left( M\left( \Omega^*_b \right)-M\left( \Omega_b \right)\right)-
2\left( M\left( \Xi^*_b \right)-M\left( \Xi'_b \right)\right)=0 \, ,
\end{equation}

\begin{equation}
M\left( \Sigma_b \right)+M\left( \Omega_b \right)-
2M\left( \Xi'_b \right)=0\, .
\end{equation}

The sign in our prediction is

\begin{equation}
M\left( \Sigma^*_b \right)-M\left( \Sigma_b \right)<
M\left( \Omega^*_b \right)-M\left( \Omega_b \right)\, .
\end{equation}

This inequality is not predicted by other recent approaches \cite{33, 34},
which predict a $\Omega_b$ splitting smaller than a $\Sigma_b$ splitting.
This suggests that the sign of the $SU(3)$ symmetry breaking gives
information about the form of the confinement potential.

We have used the $m_b/m_c=2.95$ in the Tables \ref{tab1} and \ref{tab2}
(similar to the Ref. \cite{35}).

\section{Six-quark amplitudes of the hexaquarks.}

We derive the relativistic six-quark equations in the framework of the
dispersion relation technique. We use only planar diagrams; the other
diagrams due to the rules of $1/N_c$ expansion \cite{24, 25, 26} are neglected.
The current generates a six-quark system. The correct equations for the
amplitude are obtained by taking into account all possible subamplitudes.
Then one should represent a six-particle amplitude as a sum of 15 subamplitudes:

\begin{eqnarray}
A=\sum\limits_{i<j \atop i, j=1}^6 A_{ij}\, . \end{eqnarray}

This defines the division of the diagrams into groups according to the
certain pair interaction of particles. The total amplitude can be
represented graphically as a sum of diagrams. We need to consider only
one group of diagrams and the amplitude corresponding to them, for example
$A_{12}$. We shall consider the derivation of the relativistic
generalization of the Faddeev-Yakubovsky approach.
In our case, the low-lying dibaryons are considered. We take into account the
pairwise interaction of all six quarks in the hexaquark.

For instance, we consider the state $\Delta\Lambda_b$ with $I=\frac{3}{2}$, $J^P=2^+$
and quark content $uuuudb$. The set of diagrams associated with the amplitude
$A_{12}$ can further be broken down into eight groups corresponding to subamplitudes:
$A_1^{1^{uu}}$, $A_1^{0^{ud}}$, $A_1^{0^{ub}}$, $A_1^{0^{db}}$, $A_2^{1^{uu}0^{ud}}$,
$A_2^{1^{uu}0^{ub}}$, $A_2^{1^{uu}0^{db}}$, $A_3^{1^{uu}1^{uu}0^{db}}$.

The amplitude $A_1^{1^{uu}}(s,s_{12345},s_{1234},s_{123},s_{12})$ consists
of the three color sub-structures: the diquark $1^{uu}$ in the color state
$\bar 3_c$, the quarks 3, 4 in the color state
$3_c \times 3_c=\bar 3_c +6_c$, and the quarks 5, 6 in the color state
$3_c \times 3_c=\bar 3_c +6_c$. Then we consider the total color singlet:
$\bar 3_c \times \bar 3_c \times \bar 3_c=1_c +8_c +8_c +10_c^*$.
The dibaryon amplitude
$A_2^{1^{uu}0^{ud}}(s,s_{12345},s_{1234},s_{12},s_{34})$ contains the
following sub-structures: the two diquark $1^{uu}$ and $0^{ud}$ in the color state
$\bar 3_c$ and the two quarks in the color state $3_c$. Then the dibaryon
amplitude is the total color singlet.
The amplitude $A_3^{1^{uu}1^{uu}0^{db}}(s,s_{12345},s_{12},s_{34},s_{56})$
consists of the three diquark structures in the color state $\bar 3_c$.
Therefore the total color singlet can be constructed.
For the others amplitudes color singlet also can be found.

The system of graphical equations (see for example equation for the amplitude
$A_2^{1^{uu}0^{ud}}$ for the state $\Delta\Lambda_b$ with $I=\frac{3}{2}$ and
$J^P=2^+$ at the Fig. 1) is determined by the subamplitudes
using the self-consistent method. The coefficients are determined by the
permutation of quarks.

In order to represent the subamplitudes $A_1^{1^{uu}}$, $A_1^{0^{ud}}$,
$A_1^{0^{ub}}$, $A_1^{0^{db}}$, $A_2^{1^{uu}0^{ud}}$, $A_2^{1^{uu}0^{ub}}$,
$A_2^{1^{uu}0^{db}}$, $A_3^{1^{uu}1^{uu}0^{db}}$ in the form of a
dispersion relation, it is necessary to define the amplitude of $qq$
and $qQ$ interactions. We use the results of our relativistic quark
model \cite{29} and write down the pair quark amplitudes in the form:

\begin{equation}
a_n(s_{ik})=\frac{G^2_n(s_{ik})}
{1-B_n(s_{ik})} \, ,\end{equation}

\begin{equation}
B_n(s_{ik})=\int\limits_{(m_i+m_k)^2}^{\frac{(m_i+m_k)^2\Lambda}{4}}
\hskip2mm \frac{ds'_{ik}}{\pi}\frac{\rho_n(s'_{ik})G^2_n(s'_{ik})}
{s'_{ik}-s_{ik}} \, ,\end{equation}

\begin{eqnarray}
\rho_n (s_{ik},J^{PC})&=&\left(\alpha(n,J^{PC}) \frac{s_{ik}}{(m_i+m_k)^2}
+\beta(n,J^{PC})+\delta(n,J^{PC}) \frac{(m_i-m_k)^2}{s_{ik}}\right)
\nonumber\\
&&\nonumber\\
&\times & \frac{\sqrt{(s_{ik}-(m_i+m_k)^2)(s_{ik}-(m_i-m_k)^2)}}
{s_{ik}}\, .
\end{eqnarray}

The coefficients $\alpha(n,J^{PC})$, $\beta(n,J^{PC})$ and
$\delta(n,J^{PC})$ are given in Table \ref{tab8}.
Here $n=1$ coresponds to $qq$ and $qQ$-pairs with $J^P=0^+$, $n=2$ corresponds
to $qq$ and $qQ$-pairs with $J^P=1^+$.

The coupled integral equations correspond to Fig. 1 can be described
similar to \cite{36}. Then we can go from the integration of the cosine
of the angles $dz_i$ to the integration over the subenergies.

Let us extract two- and three-particle singularities in the amplitudes
$A_1^{1^{uu}}$, $A_1^{0^{ud}}$, $A_1^{0^{ub}}$, $A_1^{0^{db}}$,
$A_2^{1^{uu}0^{ud}}$, $A_2^{1^{uu}0^{ub}}$, $A_2^{1^{uu}0^{db}}$,
$A_3^{1^{uu}1^{uu}0^{db}}$:

\begin{eqnarray}
A_1^{1^{uu}}(s,s_{12345},s_{1234},s_{123},s_{12})&=
&\frac{\alpha_1^{1^{uu}} (s,s_{12345},s_{1234},s_{123},s_{12})
B_{1^{uu}}(s_{12})}{[1-B_{1^{uu}}(s_{12})]} \, ,\\
&&\nonumber\\
A_1^{0^{ud}}(s,s_{12345},s_{1234},s_{123},s_{12})&=
&\frac{\alpha_1^{0^{ud}} (s,s_{12345},s_{1234},s_{123},s_{12})
B_{0^{ud}}(s_{12})}{[1-B_{0^{ud}}(s_{12})]} \, ,\\
&&\nonumber\\
A_1^{0^{ub}}(s,s_{12345},s_{1234},s_{123},s_{12})&=
&\frac{\alpha_1^{0^{ub}} (s,s_{12345},s_{1234},s_{123},s_{12})
B_{0^{ub}}(s_{12})}{[1-B_{0^{ub}}(s_{12})]} \, ,\\
&&\nonumber\\
A_1^{0^{db}}(s,s_{12345},s_{1234},s_{123},s_{12})&=
&\frac{\alpha_1^{0^{db}} (s,s_{12345},s_{1234},s_{123},s_{12})
B_{0^{db}}(s_{12})}{[1-B_{0^{db}}(s_{12})]} \, ,\\
&&\nonumber\\
A_2^{1^{uu}0^{ud}}(s,s_{12345},s_{1234},s_{12},s_{34})&=
&\frac{\alpha_2^{1^{uu}0^{ud}} (s,s_{12345},s_{1234},s_{12},s_{34})
B_{1^{uu}}(s_{12})B_{0^{ud}}(s_{34})}{[1-B_{1^{uu}}(s_{12})]
[1-B_{0^{ud}}(s_{34})]} \, , \\
&&\nonumber\\
A_2^{1^{uu}0^{ub}}(s,s_{12345},s_{1234},s_{12},s_{34})&=
&\frac{\alpha_2^{1^{uu}0^{ub}} (s,s_{12345},s_{1234},s_{12},s_{34})
B_{1^{uu}}(s_{12})B_{0^{ub}}(s_{34})}{[1-B_{1^{uu}}(s_{12})]
[1-B_{0^{ub}}(s_{34})]} \, , \\
&&\nonumber\\
A_2^{1^{uu}0^{db}}(s,s_{12345},s_{1234},s_{12},s_{34})&=
&\frac{\alpha_2^{1^{uu}0^{db}} (s,s_{12345},s_{1234},s_{12},s_{34})
B_{1^{uu}}(s_{12})B_{0^{db}}(s_{34})}{[1-B_{1^{uu}}(s_{12})]
[1-B_{0^{db}}(s_{34})]} \, , \\
&&\nonumber\\
A_3^{1^{uu}1^{uu}0^{db}}(s,s_{12345},s_{12},s_{34},s_{56})&=
&\frac{\alpha_3^{1^{uu}1^{uu}0^{db}} (s,s_{12345},s_{12},s_{34},s_{56})
B_{1^{uu}}(s_{12})B_{1^{uu}}(s_{34}) B_{0^{db}}(s_{56})}
{[1- B_{1^{uu}}(s_{12})] [1- B_{1^{uu}}(s_{34})][1- B_{0^{db}}(s_{56})]}
 \, . \nonumber\\
&&
\end{eqnarray}

We used the classification of singularities, which was proposed in
paper \cite{31}. Using this classification, one defines the reduced
amplitudes $\alpha_1$, $\alpha_2$, $\alpha_3$ as well as the $B$-functions
in the middle point of physical region of Dalitz-plot at the point $s_0$.

Such choice of point $s_0$ allows us to replace integral equations
($\Delta\Lambda_b$, $I=\frac{3}{2}$, $J^P=2^+$)
by the algebraic equations (\ref{24}) -- (\ref{31}):

\begin{eqnarray}
\label{24}
\alpha_1^{1^{uu}}&=&\lambda+4\, \alpha_1^{1^{uu}} I_1(1^{uu}1^{uu})
+2\, \alpha_1^{0^{ud}} I_1(1^{uu}0^{ud})
+2\, \alpha_1^{0^{ub}} I_1(1^{uu}0^{ub})
+4\, \alpha_2^{1^{uu}0^{ud}} I_2(1^{uu}1^{uu}0^{ud})
\nonumber\\
&&\nonumber\\
&+&4\, \alpha_2^{1^{uu}0^{ub}} I_2(1^{uu}1^{uu}0^{ub}) \, ,
\\
&&\nonumber\\
\label{25}
\alpha_1^{0^{ud}}&=&\lambda+3\, \alpha_1^{1^{uu}} I_1(0^{ud}1^{uu})
+3\, \alpha_1^{0^{ud}} I_1(0^{ud}0^{ud})
+\alpha_1^{0^{ub}} I_1(0^{ud}0^{ub})
+\alpha_1^{0^{db}} I_1(0^{ud}0^{db})
\nonumber\\
&&\nonumber\\
&+&6\, \alpha_2^{1^{uu}0^{ud}} I_2(0^{ud}1^{uu}0^{ud})
+3\, \alpha_2^{1^{uu}0^{db}} I_2(0^{ud}1^{uu}0^{db}) \, ,
\\
&&\nonumber\\
\label{26}
\alpha_1^{0^{ub}}&=&\lambda+3\, \alpha_1^{1^{uu}} I_1(0^{ub}1^{uu})
+\alpha_1^{0^{ud}} I_1(0^{ub}0^{ud})
+3\, \alpha_1^{0^{ub}} I_1(0^{ub}0^{ub})
+\alpha_1^{0^{db}} I_1(0^{ub}0^{db})
\nonumber\\
&&\nonumber\\
&+&6\, \alpha_2^{1^{uu}0^{ub}} I_2(0^{ub}1^{uu}0^{ub})
+3\, \alpha_2^{1^{uu}0^{db}} I_2(0^{ub}1^{uu}0^{db}) \, ,
\\
&&\nonumber\\
\label{27}
\alpha_1^{0^{db}}&=&\lambda+4\, \alpha_1^{0^{ud}} I_1(0^{db}0^{ud})
+4\, \alpha_1^{0^{ub}} I_1(0^{db}0^{ub}) \, ,
\\
&&\nonumber\\
\label{28}
\alpha_2^{1^{uu}0^{ud}}&=&\lambda
+\alpha_1^{1^{uu}} (2\, I_3(1^{uu}0^{ud}1^{uu})+2\, I_4(1^{uu}0^{ud}1^{uu}))
+2\, \alpha_1^{0^{ud}} I_3(1^{uu}0^{ud}0^{ud})
+\alpha_1^{0^{ub}} I_4(0^{ud}1^{uu}0^{ub})
\nonumber\\
&&\nonumber\\
&+&\alpha_1^{0^{db}} I_4(0^{ud}1^{uu}0^{db})
+2\, \alpha_2^{1^{uu}0^{ud}} I_7(0^{ud}1^{uu}0^{ud}1^{uu})
+\alpha_2^{1^{uu}0^{ub}} (2\, I_5(1^{uu}0^{ud}1^{uu}0^{ub})
\nonumber\\
&&\nonumber\\
&+&2\, I_6(1^{uu}0^{ud}1^{uu}0^{ub}))
+\alpha_2^{1^{uu}0^{db}} (I_5(0^{ud}1^{uu}1^{uu}0^{db})
+2\, I_6(1^{uu}0^{ud}1^{uu}0^{db})+2\, I_7(1^{uu}0^{ud}1^{uu}0^{db}))
\nonumber\\
&&\nonumber\\
&+&2\, \alpha_3^{1^{uu}1^{uu}0^{db}} I_8(1^{uu}0^{ud}1^{uu}1^{uu}0^{db}) \, ,
\\
&&\nonumber\\
\label{29}
\alpha_2^{1^{uu}0^{ub}}&=&\lambda
+\alpha_1^{1^{uu}} (2\, I_3(1^{uu}0^{ub}1^{uu})+2\, I_4(1^{uu}0^{ub}1^{uu}))
+\alpha_1^{0^{ud}} I_4(0^{ub}1^{uu}0^{ud})
+2\, \alpha_1^{0^{ub}} I_3(1^{uu}0^{ub}0^{ub})
\nonumber\\
&&\nonumber\\
&+&\alpha_1^{0^{db}} I_4(0^{ub}1^{uu}0^{db})
+\alpha_2^{1^{uu}0^{ud}} (2\, I_5(1^{uu}0^{ub}1^{uu}0^{ud})
+2\, I_6(1^{uu}0^{ub}1^{uu}0^{ud}))
\nonumber\\
&&\nonumber\\
&+&2\, \alpha_2^{1^{uu}0^{ub}} I_7(0^{ub}1^{uu}0^{ub}1^{uu}))
+\alpha_2^{1^{uu}0^{db}} (I_5(0^{ub}1^{uu}1^{uu}0^{db})
+2\, I_6(1^{uu}0^{ub}1^{uu}0^{db})
\nonumber\\
&&\nonumber\\
&+&2\, I_7(1^{uu}0^{ub}1^{uu}0^{db}))
+2\, \alpha_3^{1^{uu}1^{uu}0^{db}} I_8(1^{uu}0^{ub}1^{uu}1^{uu}0^{db}) \, ,
\\
&&\nonumber\\
\label{30}
\alpha_2^{1^{uu}0^{db}}&=&\lambda
+4\, \alpha_1^{1^{uu}} I_4(1^{uu}0^{db}1^{uu})
+\alpha_1^{0^{ud}} (2\, I_3(1^{uu}0^{db}0^{ud})+2\, I_4(0^{db}1^{uu}0^{ud}))
+\alpha_1^{0^{ub}} (2\, I_3(1^{uu}0^{db}0^{ub})
\nonumber\\
&&\nonumber\\
&+&2\, I_4(0^{db}1^{uu}0^{ub}))
+\alpha_2^{1^{uu}0^{ud}} (4\, I_6(1^{uu}0^{db}1^{uu}0^{ud})
+4\, I_7(0^{db}1^{uu}0^{ud}1^{uu}))
\nonumber\\
&&\nonumber\\
&+&\alpha_2^{1^{uu}0^{ub}} (4\, I_6(1^{uu}0^{db}1^{uu}0^{ub})
+4\, I_7(0^{db}1^{uu}0^{ub}1^{uu})) \, ,
\\
&&\nonumber\\
\label{31}
\alpha_3^{1^{uu}1^{uu}0^{db}}&=&\lambda
+4\, \alpha_1^{1^{uu}} I_9(1^{uu}1^{uu}0^{db}1^{uu})
+4\, \alpha_1^{0^{ud}} I_9(1^{uu}0^{db}1^{uu}0^{ud})
+4\, \alpha_1^{0^{ub}} I_9(1^{uu}0^{db}1^{uu}0^{ub})
\nonumber\\
&&\nonumber\\
&+&8\, \alpha_2^{1^{uu}0^{ud}} I_{10}(1^{uu}1^{uu}0^{db}1^{uu}0^{ud})
+8\, \alpha_2^{1^{uu}0^{ub}} I_{10}(1^{uu}1^{uu}0^{db}1^{uu}0^{ub}) \, ,
\end{eqnarray}

\noindent
where $\lambda_i$ are the current constants. We used the functions
$I_1$, $I_2$, $I_3$, $I_4$, $I_5$, $I_6$, $I_7$, $I_8$, $I_9$, $I_{10}$:

\begin{eqnarray}
I_1(ij)&=&\frac{B_j(s_0^{13})}{B_i(s_0^{12})}
\int\limits_{(m_1+m_2)^2}^{\frac{(m_1+m_2)^2\Lambda_i}{4}}
\frac{ds'_{12}}{\pi}\frac{G_i^2(s_0^{12})\rho_i(s'_{12})}
{s'_{12}-s_0^{12}} \int\limits_{-1}^{+1} \frac{dz_1(1)}{2}
\frac{1}{1-B_j (s'_{13})}\, , \\
&&\nonumber\\
I_2(ijk)&=&\frac{B_j(s_0^{13}) B_k(s_0^{24})}{B_i(s_0^{12})}
\int\limits_{(m_1+m_2)^2}^{\frac{(m_1+m_2)^2\Lambda_i}{4}}
\frac{ds'_{12}}{\pi}\frac{G_i^2(s_0^{12})\rho_i(s'_{12})}
{s'_{12}-s_0^{12}}
\frac{1}{2\pi}\int\limits_{-1}^{+1}\frac{dz_1(2)}{2}
\int\limits_{-1}^{+1} \frac{dz_2(2)}{2}\nonumber\\
&&\nonumber\\
&\times&
\int\limits_{z_3(2)^-}^{z_3(2)^+} dz_3(2)
\frac{1}{\sqrt{1-z_1^2(2)-z_2^2(2)-z_3^2(2)+2z_1(2) z_2(2) z_3(2)}}
\nonumber\\
&&\nonumber\\
&\times& \frac{1}{1-B_j (s'_{13})} \frac{1}{1-B_k (s'_{24})}
 \, , \\
&&\nonumber\\
I_3(ijk)&=&\frac{B_k(s_0^{23})}{B_i(s_0^{12}) B_j(s_0^{34})}
\int\limits_{(m_1+m_2)^2}^{\frac{(m_1+m_2)^2\Lambda_i}{4}}
\frac{ds'_{12}}{\pi}\frac{G_i^2(s_0^{12})\rho_i(s'_{12})}
{s'_{12}-s_0^{12}}\nonumber\\
&&\nonumber\\
&\times&\int\limits_{(m_3+m_4)^2}^{\frac{(m_3+m_4)^2\Lambda_j}{4}}
\frac{ds'_{34}}{\pi}\frac{G_j^2(s_0^{34})\rho_j(s'_{34})}
{s'_{34}-s_0^{34}}
\int\limits_{-1}^{+1} \frac{dz_1(3)}{2} \int\limits_{-1}^{+1}
\frac{dz_2(3)}{2} \frac{1}{1-B_k (s'_{23})} \, , \\
&&\nonumber\\
I_4(ijk)&=&I_1(ik) \, , \\
&&\nonumber\\
I_5(ijkl)&=&I_2(ikl) \, , \\
&&\nonumber\\
I_6(ijkl)&=&I_1(ik) \cdot I_1(jl)
 \, , \\
&&\nonumber\\
I_7(ijkl)&=&\frac{B_k(s_0^{23})B_l(s_0^{45})}{B_i(s_0^{12}) B_j(s_0^{34})}
\int\limits_{(m_1+m_2)^2}^{\frac{(m_1+m_2)^2\Lambda_i}{4}}
\frac{ds'_{12}}{\pi}\frac{G_i^2(s_0^{12})\rho_i(s'_{12})}
{s'_{12}-s_0^{12}}\nonumber\\
&&\nonumber\\
&\times&\int\limits_{(m_3+m_4)^2}^{\frac{(m_3+m_4)^2\Lambda_j}{4}}
\frac{ds'_{34}}{\pi}\frac{G_j^2(s_0^{34})\rho_j(s'_{34})}
{s'_{34}-s_{34}}
\frac{1}{2\pi}\int\limits_{-1}^{+1}\frac{dz_1(7)}{2}
\int\limits_{-1}^{+1} \frac{dz_2(7)}{2}
\int\limits_{-1}^{+1} \frac{dz_3(7)}{2}
\nonumber\\
&&\nonumber\\
&\times&
\int\limits_{z_4(7)^-}^{z_4(7)^+} dz_4(7)
\frac{1}{\sqrt{1-z_1^2(7)-z_3^2(7)-z_4^2(7)+2z_1(7) z_3(7) z_4(7)}}
\nonumber\\
&&\nonumber\\
&\times& \frac{1}{1-B_k (s'_{23})} \frac{1}{1-B_l (s'_{45})}
 \, , \\
&&\nonumber\\
I_8(ijklm)&=&\frac{B_k(s_0^{15})B_l(s_0^{23})B_m(s_0^{46})}
{B_i(s_0^{12}) B_j(s_0^{34})}
\int\limits_{(m_1+m_2)^2}^{\frac{(m_1+m_2)^2\Lambda_i}{4}}
\frac{ds'_{12}}{\pi}\frac{G_i^2(s_0^{12})\rho_i(s'_{12})}
{s'_{12}-s_0^{12}}\nonumber\\
&&\nonumber\\
&\times&\int\limits_{(m_3+m_4)^2}^{\frac{(m_3+m_4)^2\Lambda_j}{4}}
\frac{ds'_{34}}{\pi}\frac{G_j^2(s_0^{34})\rho_j(s'_{34})}
{s'_{34}-s_0^{34}}\nonumber\\
&&\nonumber\\
&\times&\frac{1}{(2\pi)^2}\int\limits_{-1}^{+1}\frac{dz_1(8)}{2}
\int\limits_{-1}^{+1} \frac{dz_2(8)}{2}
\int\limits_{-1}^{+1} \frac{dz_3(8)}{2}
\int\limits_{z_4(8)^-}^{z_4(8)^+} dz_4(8)
\int\limits_{-1}^{+1} \frac{dz_5(8)}{2}
\int\limits_{z_6(8)^-}^{z_6(8)^+} dz_6(8)
\nonumber\\
&&\nonumber\\
&\times&
\frac{1}{\sqrt{1-z_1^2(8)-z_3^2(8)-z_4^2(8)+2z_1(8) z_3(8) z_4(8)}}
\nonumber\\
&&\nonumber\\
&\times&
\frac{1}{\sqrt{1-z_2^2(8)-z_5^2(8)-z_6^2(8)+2z_2(8) z_5(8) z_6(8)}}
\nonumber\\
&&\nonumber\\
&\times& \frac{1}{1-B_k (s'_{15})} \frac{1}{1-B_l (s'_{23})}
\frac{1}{1-B_m (s'_{46})}
 \, , \\
&&\nonumber\\
I_9(ijkl)&=&I_3(ijl) \, , \\
&&\nonumber\\
I_{10}(ijklm)&=
&\frac{B_l(s_0^{23})B_m(s_0^{45})}
{B_i(s_0^{12}) B_j(s_0^{34}) B_k(s_0^{56})}
\int\limits_{(m_1+m_2)^2}^{\frac{(m_1+m_2)^2\Lambda_i}{4}}
\frac{ds'_{12}}{\pi}\frac{G_i^2(s_0^{12})\rho_i(s'_{12})}{s'_{12}-s_0^{12}}
\nonumber\\
&&\nonumber\\
&\times&
\int\limits_{(m_3+m_4)^2}^{\frac{(m_3+m_4)^2\Lambda_j}{4}}
\frac{ds'_{34}}{\pi}\frac{G_j^2(s_0^{34})\rho_j(s'_{34})}
{s'_{34}-s_0^{34}}
\int\limits_{(m_5+m_6)^2}^{\frac{(m_5+m_6)^2\Lambda_k}{4}}
\frac{ds'_{56}}{\pi}\frac{G_k^2(s_0^{56})\rho_k(s'_{56})}{s'_{56}-s_0^{56}}
\nonumber\\
&&\nonumber\\
&\times&
\frac{1}{2\pi}\int\limits_{-1}^{+1}\frac{dz_1(10)}{2}
\int\limits_{-1}^{+1} \frac{dz_2(10)}{2}
\int\limits_{-1}^{+1} \frac{dz_3(10)}{2}
\int\limits_{-1}^{+1} \frac{dz_4(10)}{2}
\int\limits_{z_5(1-)^-}^{z_5(10)^+} dz_5(10)
\nonumber\\
&&\nonumber\\
&\times&
\frac{1}{\sqrt{1-z_1^2(10)-z_4^2(10)-z_5^2(10)+2z_1(10) z_4(10) z_5(10)}}
\nonumber\\
&&\nonumber\\
&\times& \frac{1}{1-B_l (s'_{23})} \frac{1}{1-B_m (s'_{45})}
 \, ,
\end{eqnarray}

\noindent
where $i$, $j$, $k$, $l$, $m$ correspond to the diquarks with the
spin-parity $J^P=0^+, 1^+$.

The solutions of the system of equations are considered as:

\begin{equation}
\alpha_i(s)=F_i(s,\lambda_i)/D(s) \, ,\end{equation}

\noindent
where zeros of $D(s)$ determinants define the masses of bound states of
dibaryons.

\section{Calculation results.}

The model in question has three parameters of previous model \cite{36}:
gluon coupling constants $g_0=0.653$ (diquark $0^+$) and $g_1=0.292$
(diquark $1^+$), cutoff parameter $\Lambda=11$. We used to the cutoff
$\Lambda_{qb}=4.43$ and the cutoff $\Lambda_{qc}=5.18$, which are determined
by $M=7300\, MeV$ (threshold $7315\, MeV$) and $M=4100\, MeV$ (threshold
$4130\, MeV$). The experimental data is absent, therefore we use the dimensionless
parameters, which are similar to the Eq. (\ref{13}). It allows us to calculate
the mass spectra of $qqqqqQ$ states.

The quark masses of the model are $m_q=495\, MeV$, $m_c=1655\, MeV$,
$m_b=4840\, MeV$. The estimation of theoretical error on the $S$-wave
hexaquarks masses is $1\, MeV$. This results was obtained by the choice
of model parameters.

We have considered 38 dibaryons with content $qqqqqQ$, $q=u, d$, $Q=c, b$.
The masses of dibaryons with $I=\frac{1}{2}$, $\frac{3}{2}$, $\frac{5}{2}$
and spin-parity $J^P=0^+$, $1^+$, $2^+$ in the Tables \ref{tab4} and \ref{tab5}
are given. The lowest mass for the $qqqqqb$ states is $M=5700\, MeV$.
The lowest mass for the $qqqqqc$ states is $M=3475\, MeV$.

The relativistic six-body approach possesses the dynamical mixing and
allows us to calculate the contributions of the subamplitudes to the
hexaquark amplitude (Tables \ref{tab6}, \ref{tab7}). The calculated
dibaryon subamplitudes $A_2$ present the main contributions to the
hexaquark amplitude (about 80 percents).

In a strongly bound systems, which include the light quarks, where
$p/m \sim 1$, the approximation by nonrelativistic kinematics and dynamics
is not justified.

In our paper, the relativistic description of three-particles amplitudes
of bottom baryons are considered. We take into account the $u$, $d$,
$c$, $b$ quarks. The mass spectrum of $S$-wave bottom baryons with one,
two, and three $b$ quarks is considered. We use only two new parameters
for the calculation of $23$ baryon masses. The other model parameters
in the our papers \cite{21, 22, 23} are given.

We have predicted the masses of baryons containing $b$ quarks using the
coupled-channel formalism. We believe that the prediction for the $S$-wave
bottom baryons based on the relativistic kinematics and dynamics allows as
to take into account the relativistic corrections. In our consideration, the
bottom baryon masses are heavier than the masses in the other quark models
\cite{37, 38}.

Our model is confined to the quark-antiquark pair production on account of
the phase space restriction. Here $m_q$ is the "mass" of the constituent
quark. Therefore the production of new quark-antiquark pair is absent
for the low-lying hadrons.

We manage with quark as with real particles. However, in the soft region,
the quark diagrams should be treated as spectral integral over quark masses
with the spectral density $\rho(m^2)$: the integration over quark masses
in the amplitudes puts away the quark singularities and introduces the
hadron ones. One can believe that the approximation: $\rho(m^2)\to \delta (m^2-m_q^2)$
could be possible for the low-lying hadrons.

The authors of Ref. \cite{38} calculated the energies of the baryon-baryon
threshold as a function of the flavor-symmetry breaking parameter
$\delta=1-\frac{m_u}{m_s}$. The binding energy is obtained and the possible
candidates for stability under strong interactions is commented on.
In the recent papers \cite{39, 40}, developed for the baryon-baryon
interactions in lattice QCD, the flavor-singlet H dibaryon is studied.
The results of the lattice QCD calculations presented the first clear evidence
for a bound state of two baryon directly from QCD.

\begin{acknowledgments}
S.M. Gerasyuta is indebted to T. Barnes, C.-Y. Wong for useful discussions.
This research was supported in part by the Russian Ministry of Education
under Grant 2.1.1.68.26.
\end{acknowledgments}

\begin{table}
\caption{Bottom baryon masses of multiplet $\frac{1}{2}^+$.
Parameters of model: quark masses $m_{u,d}=495\, MeV$, $m_s=770\, MeV$,
$m_c=1655\, MeV$, $m_b=4840\, MeV$; cutoff parameters: $\lambda_q=10.7$
($q=u, d, s$), $\lambda_c=6.5$, $\lambda_b=5.4$; gluon coupling constants:
$g_0=0.70$, $g_1=0.55$ with $J^p=0^+$ and $1^+$, $g_c=0.857$, $g_b=1.03$.}
\label{tab1}
\begin{tabular}{|c|c|c|}
\hline
Baryon & Mass ($GeV$) & Mass ($GeV$) (exp.)\\
\hline
$\Sigma_b$ & $5.808$ & $5.808$ \\
\hline
$\Lambda_b$ & $5.624$ & $5.624$ \\
\hline
$\Xi_{sb}^A$ & $5.761$ & $5.774$, $5.793$ \\
\hline
$\Xi_{sb}^S$ & $6.007$ & -- \\
\hline
$\Omega_{ssb}$ & $6.120$ & -- \\
\hline
$\Xi_{cb}^A$ & $6.789$ & -- \\
\hline
$\Xi_{cb}^S$ & $6.818$ & -- \\
\hline
$\Lambda_{scb}^A$ & $6.798$ & -- \\
\hline
$\Lambda_{scb}^S$ & $6.836$ & -- \\
\hline
$\Omega_{ccb}$ & $7.943$ & -- \\
\hline
$\Xi_{bb}$ & $10.045$ & -- \\
\hline
$\Omega_{sbb}$ & $9.999$ & -- \\
\hline
$\Omega_{cbb}$ & $11.089$ & -- \\
\hline
\end{tabular}
\end{table}

\begin{table}
\caption{Bottom baryon masses of multiplet $\frac{3}{2}^+$.
Parameters of model: quark masses $m_{u,d}=495\, MeV$, $m_s=770\, MeV$,
$m_c=1655\, MeV$, $m_b=4840\, MeV$; cutoff parameters: $\lambda_q=10.7$
($q=u, d, s$), $\lambda_c=6.5$, $\lambda_b=5.4$; gluon coupling constants:
$g_0=0.70$, $g_1=0.55$ with $J^p=0^+$ and $1^+$, $g_c=0.857$, $g_b=1.03$.}
\label{tab2}
\begin{tabular}{|c|c|c|}
\hline
Baryon & Mass ($GeV$) & Mass ($GeV$) (exp.)\\
\hline
$\Sigma_b$ & $5.829$ & $5.829$ \\
\hline
$\Xi_{sb}$ & $6.066$ & -- \\
\hline
$\Omega_{ssb}$ & $6.220$ & -- \\
\hline
$\Xi_{cb}$ & $6.863$ & -- \\
\hline
$\Omega_{scb}$ & $6.914$ & -- \\
\hline
$\Omega_{ccb}$ & $7.973$ & -- \\
\hline
$\Xi_{bb}$ & $10.104$ & -- \\
\hline
$\Omega_{sbb}$ & $10.126$ & -- \\
\hline
$\Omega_{cbb}$ & $11.123$ & -- \\
\hline
$\Omega_{bbb}$ & $14.197$ & -- \\
\hline
\end{tabular}
\end{table}

\begin{table}
\caption{Coefficients of Ghew-Mandelstam functions.}
\label{tab3}
\begin{tabular}{|c|c|c|c|}
\hline
 &$\alpha_J$&$\beta_J$&$\delta_J$\\
\hline
 & & & \\
$1^+$&$\frac{1}{3}$&$\frac{4m_i m_k}{3(m_i+m_k)^2}-\frac{1}{6}$
&$-\frac{1}{6}(m_i-m_k)^2$\\
 & & & \\
$0^+$&$\frac{1}{2}$&$-\frac{1}{2}\frac{(m_i-m_k)^2}{(m_i+m_k)^2}$&0\\
 & & & \\
\hline
\end{tabular}
\end{table}

\begin{table}
\caption{S-wave charmed dibaryon masses. Parameters of model: cutoff
$\Lambda=11.0$ and $\Lambda_{qc}=5.18$, gluon coupling constants
$g_0=0.653$ and $g_1=0.292$. Quark masses $m_q=495\, MeV$ and
$m_c=1655\, MeV$.}
\label{tab4}
\begin{tabular}{|c|c|c|c|}
\hline
$I$ & $J$ & Dibaryons (quark content) & Mass (MeV) \\
\hline
$\frac{5}{2}$ & 0    & $\Delta\Sigma^*_c$ ($uuu\,\, uuc$) & 4100 \\
              & 1, 2 & $\Delta\Sigma_c$, $\Delta\Sigma^*_c$ ($uuu\,\, uuc$)
                                                          & 4100 \\
$\frac{3}{2}$ & 0    & $\Delta\Sigma^*_c$ ($uuu\,\, udc+uud\,\, uuc$)
                                                          & 3570 \\
              &      & $N\Sigma_c$ ($uud\,\, uuc$)        & 3699 \\
              & 1    & $\Delta\Sigma_c$, $\Delta\Sigma^*_c$
                              ($uuu\,\, udc+uud\,\, uuc$) & 3570 \\
              &      & $\Delta\Lambda_c$ ($uuu\,\, udc$)  & 3920 \\
              &      & $N\Sigma_c$, $N\Sigma^*_c$
                                          ($uud\,\, uuc$) & 3699 \\
              & 2    & $\Delta\Sigma_c$, $\Delta\Sigma^*_c$
                              ($uuu\,\, udc+uud\,\, uuc$) & 3746 \\
              &      & $\Delta\Lambda_c$ ($uuu\,\, udc$)  & 3920 \\
              &      & $N\Sigma^*_c$ ($uud\,\, uuc$)      & 3902 \\
$\frac{1}{2}$ & 0   & $\Delta\Sigma^*_c$ ($uuu\,\, ddc+uud\,\, udc+udd\,\,
                                                    uuc$) & 3475 \\
              &     & $N\Lambda_c$ ($uud\,\, udc$)        & 3629 \\
              &     & $N\Sigma_c$ ($uud\,\, udc+udd\,\, uuc$)
                                                          & 3480 \\
              & 1   & $\Delta\Sigma_c$, $\Delta\Sigma^*_c$ ($uuu\,\,
                            ddc+uud\,\, udc+udd\,\, uuc$) & 3475 \\
              &     & $N\Lambda_c$, $\Delta\Lambda_c$ ($uud\,\, udc$)
                                                          & 3629 \\
              &     & $N\Sigma_c$, $N\Sigma^*_c$ ($uud\,\, udc+udd\,\,
                                                    uuc$) & 3480 \\
              & 2   & $\Delta\Sigma_c$, $\Delta\Sigma^*_c$ ($uuu\,\,
                            ddc+uud\,\, udc+udd\,\, uuc$) & 3863 \\
              &     & $\Delta\Lambda_c$ ($uud\,\, udc$)   & 3937 \\
              &     & $N\Sigma^*_c$ ($uud\,\, udc+udd\,\, uuc$)
                                                          & 3870 \\
\hline
\end{tabular}
\end{table}

\begin{table}
\caption{S-wave bottom dibaryon masses. Parameters of model: cutoff
$\Lambda=11.0$ and $\Lambda_{qb}=4.43$, gluon coupling constants
$g_0=0.653$ and $g_1=0.292$. Quark masses $m_q=495\, MeV$ and
$m_b=4840\, MeV$.}
\label{tab5}
\begin{tabular}{|c|c|c|c|}
\hline
$I$ & $J$ & Dibaryons (quark content) & Mass (MeV) \\
\hline
$\frac{5}{2}$ & 0    & $\Delta\Sigma^*_b$ ($uuu\,\, uub$) & 7300 \\
              & 1, 2 & $\Delta\Sigma_b$, $\Delta\Sigma^*_b$ ($uuu\,\, uub$)
                                                          & 7300 \\
$\frac{3}{2}$ & 0    & $\Delta\Sigma^*_b$ ($uuu\,\, udb+uud\,\, uub$)
                                                          & 5988 \\
              &      & $N\Sigma_b$ ($uud\,\, uub$)        & 6285 \\
              & 1    & $\Delta\Sigma_b$, $\Delta\Sigma^*_b$
                              ($uuu\,\, udb+uud\,\, uub$) & 5988 \\
              &      & $\Delta\Lambda_b$ ($uuu\,\, udb$)  & 6926 \\
              &      & $N\Sigma_b$, $N\Sigma^*_b$
                                          ($uud\,\, uub$) & 6285 \\
              & 2    & $\Delta\Sigma_b$, $\Delta\Sigma^*_b$
                              ($uuu\,\, udb+uud\,\, uub$) & 6450 \\
              &      & $\Delta\Lambda_b$ ($uuu\,\, udb$)  & 6926 \\
              &      & $N\Sigma^*_b$ ($uud\,\, uub$)      & 6800 \\
$\frac{1}{2}$ & 0   & $\Delta\Sigma^*_b$ ($uuu\,\, ddb+uud\,\, udb+udd\,\,
                                                    uub$) & 5700 \\
              &     & $N\Lambda_b$ ($uud\,\, udb$)        & 6142 \\
              &     & $N\Sigma_b$ ($uud\,\, udb+udd\,\, uub$)
                                                          & 5723 \\
              & 1   & $\Delta\Sigma_b$, $\Delta\Sigma^*_b$ ($uuu\,\,
                            ddb+uud\,\, udb+udd\,\, uub$) & 5700 \\
              &     & $N\Lambda_b$, $\Delta\Lambda_b$ ($uud\,\, udb$)
                                                          & 6142 \\
              &     & $N\Sigma_b$, $N\Sigma^*_b$ ($uud\,\, udb+udd\,\,
                                                    uub$) & 5723 \\
              & 2   & $\Delta\Sigma_b$, $\Delta\Sigma^*_b$ ($uuu\,\,
                            ddb+uud\,\, udb+udd\,\, uub$) & 6744 \\
              &     & $\Delta\Lambda_b$ ($uud\,\, udb$)   & 6928 \\
              &     & $N\Sigma^*_b$ ($uud\,\, udb+udd\,\, uub$)
                                                          & 6755 \\
\hline
\end{tabular}
\end{table}

\begin{table}
\caption{$\Delta\Sigma_c$, $\Delta\Sigma^*_c$ ($3475\, MeV$)
$(IJ=\frac{1}{2} 1)$.
Parameters of model: cutoff $\Lambda=11.0$ and $\Lambda_{qc}=5.18$,
gluon coupling constants $g_0=0.653$ and $g_1=0.292$. Quark masses
$m_q=495\, MeV$ and $m_c=1655\, MeV$.}
\label{tab6}
\begin{tabular}{|c|c|}
\hline
Subamplitudes & Contributions, percent \\
\hline
$A_1^{1^{uu}}$ & 4.98 \\
$A_1^{1^{dd}}$ & 4.20 \\
$A_1^{0^{ud}}$ & 3.81 \\
$A_1^{0^{uc}}$ & 0.82 \\
$A_1^{0^{dc}}$ & 0.23 \\
$A_2^{1^{uu}1^{dd}}$ & 9.58 \\
$A_2^{1^{uu}0^{ud}}$ & 17.73 \\
$A_2^{1^{uu}0^{uc}}$ & 1.17 \\
$A_2^{1^{uu}0^{dc}}$ & 1.69 \\
$A_2^{1^{dd}0^{uc}}$ & 1.43 \\
$A_2^{0^{ud}0^{ud}}$ & 46.54 \\
$A_2^{0^{ud}0^{uc}}$ & 3.23 \\
$A_2^{0^{ud}0^{dc}}$ & 3.11 \\
$A_3^{1^{uu}1^{dd}0^{uc}}$ & 0.43 \\
$A_3^{1^{uu}0^{ud}0^{dc}}$ & 1.06 \\
\hline
$\sum A_1$ & 14.05 \\
$\sum A_2$ & 84.46 \\
$\sum A_3$ & 1.49 \\
\hline
\end{tabular}
\end{table}

\begin{table}
\caption{$\Delta\Sigma_b$, $\Delta\Sigma^*_b$ ($5700\, MeV$)
$(IJ=\frac{1}{2} 1)$.
Parameters of model: cutoff $\Lambda=11.0$ and $\Lambda_{qb}=4.43$,
gluon coupling constants $g_0=0.653$ and $g_1=0.292$. Quark masses
$m_q=495\, MeV$ and $m_b=4840\, MeV$.}
\label{tab7}
\begin{tabular}{|c|c|}
\hline
Subamplitudes & Contributions, percent \\
\hline
$A_1^{1^{uu}}$ & 5.79 \\
$A_1^{1^{dd}}$ & 4.89 \\
$A_1^{0^{ud}}$ & 2.67 \\
$A_1^{0^{ub}}$ & 0.11 \\
$A_1^{0^{db}}$ & 0.02 \\
$A_2^{1^{uu}1^{dd}}$ & 11.04 \\
$A_2^{1^{uu}0^{ud}}$ & 19.82 \\
$A_2^{1^{uu}0^{ub}}$ & 0.13 \\
$A_2^{1^{uu}0^{db}}$ & 0.21 \\
$A_2^{1^{dd}0^{ub}}$ & 0.16 \\
$A_2^{0^{ud}0^{ud}}$ & 54.15 \\
$A_2^{0^{ud}0^{ub}}$ & 0.42 \\
$A_2^{0^{ud}0^{db}}$ & 0.39 \\
$A_3^{1^{uu}1^{dd}0^{ub}}$ & 0.06 \\
$A_3^{1^{uu}0^{ud}0^{db}}$ & 0.14 \\
\hline
$\sum A_1$ & 13.47 \\
$\sum A_2$ & 86.33 \\
$\sum A_3$ & 0.20 \\
\hline
\end{tabular}
\end{table}

\begin{table}
\caption{Vertex functions and Chew-Mandelstam coefficients.}\label{tab8}
\begin{tabular}{|c|c|c|c|c|}
\hline
$i$ & $G_i^2(s_{kl})$ & $\alpha_i$ & $\beta_i$ & $\delta_i$ \\
\hline
& & & & \\
$0^+$ & $\frac{4g}{3}-\frac{8gm_{kl}^2}{(3s_{kl})}$
& $\frac{1}{2}$ & $-\frac{1}{2}\frac{(m_k-m_l)^2}{(m_k+m_l)^2}$ & $0$ \\
& & & & \\
$1^+$ & $\frac{2g}{3}$ & $\frac{1}{3}$
& $\frac{4m_k m_l}{3(m_k+m_l)^2}-\frac{1}{6}$
& $-\frac{1}{6}\frac{(m_k-m_l)^2}{(m_k+m_l)^2}$ \\
& & & & \\
\hline
\end{tabular}
\end{table}

\newpage

\begin{picture}(600,600)
\put(0,545){\line(1,0){18}}
\put(0,547){\line(1,0){17.5}}
\put(0,549){\line(1,0){17}}
\put(0,551){\line(1,0){17}}
\put(0,553){\line(1,0){17.5}}
\put(0,555){\line(1,0){18}}
\put(30,550){\circle{25}}
\put(19,546){\line(1,1){15}}
\put(22,541){\line(1,1){17}}
\put(27.5,538.5){\line(1,1){14}}
\put(31,563){\vector(1,1){20}}
\put(31,538){\vector(1,-1){20}}
\put(47.5,560){\circle{16}}
\put(47.5,540){\circle{16}}
\put(55,564){\vector(3,2){18}}
\put(55,536){\vector(3,-2){18}}
\put(55,564){\vector(3,-2){18}}
\put(55,536){\vector(3,2){18}}
\put(78,575){1}
\put(78,553){2}
\put(78,541){3}
\put(78,518){4}
\put(54,580){5}
\put(54,513){6}
\put(40.5,556){\small $1^{uu}$}
\put(40.5,536){\small $0^{ud}$}
\put(90,548){$=$}
\put(110,545){\line(1,0){19}}
\put(110,547){\line(1,0){21}}
\put(110,549){\line(1,0){23}}
\put(110,551){\line(1,0){23}}
\put(110,553){\line(1,0){21}}
\put(110,555){\line(1,0){19}}
\put(140,560){\circle{16}}
\put(140,540){\circle{16}}
\put(147.5,564){\vector(3,2){18}}
\put(147.5,536){\vector(3,-2){18}}
\put(147.5,564){\vector(3,-2){18}}
\put(147.5,536){\vector(3,2){18}}
\put(128,555){\vector(1,3){11}}
\put(128,545){\vector(1,-3){11}}
\put(170,575){1}
\put(170,553){2}
\put(170,541){3}
\put(170,518){4}
\put(143,586){5}
\put(143,508){6}
\put(133,557){\small $1^{uu}$}
\put(133,537){\small $0^{ud}$}
\put(183,548){$+$}
\put(199,548){2}
\put(212,545){\line(1,0){18}}
\put(212,547){\line(1,0){17.5}}
\put(212,549){\line(1,0){17}}
\put(212,551){\line(1,0){17}}
\put(212,553){\line(1,0){17.5}}
\put(212,555){\line(1,0){18}}
\put(242,550){\circle{25}}
\put(231,546){\line(1,1){15}}
\put(234,541){\line(1,1){17}}
\put(239.5,538.5){\line(1,1){14}}
\put(243,563){\vector(1,1){20}}
\put(243,538){\vector(1,-1){20}}
\put(266,580){5}
\put(266,513){6}
\put(262,550){\circle{16}}
\put(270,550){\vector(3,2){17}}
\put(270,550){\vector(3,-2){17}}
\put(247,561){\vector(1,0){40}}
\put(247,539){\vector(1,0){40}}
\put(295,560){\circle{16}}
\put(295,540){\circle{16}}
\put(303,561){\vector(3,1){20}}
\put(303,561){\vector(3,-1){20}}
\put(303,539){\vector(3,1){20}}
\put(303,539){\vector(3,-1){20}}
\put(328,570){1}
\put(328,553){2}
\put(328,541){3}
\put(328,524){4}
\put(270,566){1}
\put(270,554){\small 2}
\put(270,540){\small 3}
\put(270,528){4}
\put(255,547){\small $1^{uu}$}
\put(288,557){\small $1^{uu}$}
\put(288,537){\small $0^{ud}$}
\put(341,548){$+$}
\put(355,548){2}
\put(368,545){\line(1,0){18}}
\put(368,547){\line(1,0){17.5}}
\put(368,549){\line(1,0){17}}
\put(368,551){\line(1,0){17}}
\put(368,553){\line(1,0){17.5}}
\put(368,555){\line(1,0){18}}
\put(398,550){\circle{25}}
\put(387,546){\line(1,1){15}}
\put(390,541){\line(1,1){17}}
\put(395.5,538.5){\line(1,1){14}}
\put(399,563){\vector(1,1){20}}
\put(399,538){\vector(1,-1){20}}
\put(422,580){5}
\put(422,513){6}
\put(418,550){\circle{16}}
\put(426,550){\vector(3,2){17}}
\put(426,550){\vector(3,-2){17}}
\put(403,561){\vector(1,0){40}}
\put(403,539){\vector(1,0){40}}
\put(451,560){\circle{16}}
\put(451,540){\circle{16}}
\put(459,561){\vector(3,1){20}}
\put(459,561){\vector(3,-1){20}}
\put(459,539){\vector(3,1){20}}
\put(459,539){\vector(3,-1){20}}
\put(482,570){1}
\put(482,553){2}
\put(482,541){3}
\put(482,524){4}
\put(426,566){1}
\put(426,554){\small 2}
\put(426,540){\small 3}
\put(426,528){4}
\put(411,547){\small $0^{ud}$}
\put(444,557){\small $1^{uu}$}
\put(444,537){\small $0^{ud}$}
\put(10,448){$+$}
\put(24,448){2}
\put(37,445){\line(1,0){18}}
\put(37,447){\line(1,0){17.5}}
\put(37,449){\line(1,0){17}}
\put(37,451){\line(1,0){17}}
\put(37,453){\line(1,0){17.5}}
\put(37,455){\line(1,0){18}}
\put(67,450){\circle{25}}
\put(56,446){\line(1,1){15}}
\put(59,441){\line(1,1){17}}
\put(64.5,438.5){\line(1,1){14}}
\put(83,462){\circle{16}}
\put(88.5,468.5){\vector(1,1){15}}
\put(88.5,468.5){\vector(1,-1){18}}
\put(79,450){\vector(1,0){28}}
\put(115,450){\circle{16}}
\put(123,450){\vector(3,1){22}}
\put(123,450){\vector(3,-1){22}}
\put(99,461){1}
\put(96,439){2}
\put(139,462){1}
\put(139,430){2}
\put(105,485){5}
\put(83,437){\circle{16}}
\put(90,432){\vector(3,-1){20}}
\put(90,432){\vector(2,-3){12}}
\put(68,438){\vector(1,-3){8}}
\put(113,422){3}
\put(104,407){4}
\put(78,407){6}
\put(76,459){\small $1^{uu}$}
\put(108,447){\small $1^{uu}$}
\put(76,434){\small $0^{ud}$}
\put(175,448){$+$}
\put(200,445){\line(1,0){18}}
\put(200,447){\line(1,0){17.5}}
\put(200,449){\line(1,0){17}}
\put(200,451){\line(1,0){17}}
\put(200,453){\line(1,0){17.5}}
\put(200,455){\line(1,0){18}}
\put(230,450){\circle{25}}
\put(219,446){\line(1,1){15}}
\put(222,441){\line(1,1){17}}
\put(227.5,438.5){\line(1,1){14}}
\put(246,462){\circle{16}}
\put(251.5,468.5){\vector(1,1){15}}
\put(251.5,468.5){\vector(1,-1){18}}
\put(242,450){\vector(1,0){28}}
\put(278,450){\circle{16}}
\put(286,450){\vector(3,1){22}}
\put(286,450){\vector(3,-1){22}}
\put(262,461){1}
\put(259,439){2}
\put(302,462){1}
\put(302,430){2}
\put(268,485){5}
\put(246,437){\circle{16}}
\put(253,432){\vector(3,-1){20}}
\put(253,432){\vector(2,-3){12}}
\put(231,438){\vector(1,-3){8}}
\put(276,422){3}
\put(267,407){4}
\put(241,407){6}
\put(239,459){\small $0^{ub}$}
\put(271,447){\small $0^{ud}$}
\put(239,434){\small $1^{uu}$}
\put(340,448){$+$}
\put(368,445){\line(1,0){18}}
\put(368,447){\line(1,0){17.5}}
\put(368,449){\line(1,0){17}}
\put(368,451){\line(1,0){17}}
\put(368,453){\line(1,0){17.5}}
\put(368,455){\line(1,0){18}}
\put(398,450){\circle{25}}
\put(387,446){\line(1,1){15}}
\put(390,441){\line(1,1){17}}
\put(395.5,438.5){\line(1,1){14}}
\put(414,462){\circle{16}}
\put(419.5,468.5){\vector(1,1){15}}
\put(419.5,468.5){\vector(1,-1){18}}
\put(410,450){\vector(1,0){28}}
\put(446,450){\circle{16}}
\put(454,450){\vector(3,1){22}}
\put(454,450){\vector(3,-1){22}}
\put(430,461){1}
\put(427,439){2}
\put(470,462){1}
\put(470,430){2}
\put(436,485){5}
\put(414,437){\circle{16}}
\put(421,432){\vector(3,-1){20}}
\put(421,432){\vector(2,-3){12}}
\put(399,438){\vector(1,-3){8}}
\put(444,422){3}
\put(435,407){4}
\put(409,407){6}
\put(407,459){\small $0^{db}$}
\put(439,447){\small $0^{ud}$}
\put(407,434){\small $1^{uu}$}
\put(10,348){$+$}
\put(24,348){2}
\put(37,345){\line(1,0){18}}
\put(37,347){\line(1,0){17.5}}
\put(37,349){\line(1,0){17}}
\put(37,351){\line(1,0){17}}
\put(37,353){\line(1,0){17.5}}
\put(37,355){\line(1,0){18}}
\put(67,350){\circle{25}}
\put(56,346){\line(1,1){15}}
\put(59,341){\line(1,1){17}}
\put(64.5,338.5){\line(1,1){14}}
\put(84.5,360){\circle{16}}
\put(84.5,340){\circle{16}}
\put(89,367){\vector(1,1){18}}
\put(89,333){\vector(1,-1){18}}
\put(89,367){\vector(1,-1){17}}
\put(89,333){\vector(1,1){17}}
\put(114,350){\circle{16}}
\put(122,350){\vector(1,1){18}}
\put(122,350){\vector(1,-1){18}}
\put(67,330){\circle{16}}
\put(67,322){\vector(1,-2){9}}
\put(67,322){\vector(-1,-2){9}}
\put(98,362){1}
\put(98,331){2}
\put(143,365){1}
\put(143,327){2}
\put(111,385){5}
\put(111,310){6}
\put(79,305){3}
\put(49,305){4}
\put(77.5,357){\small $1^{uu}$}
\put(77.5,337){\small $0^{ub}$}
\put(107,347){\small $1^{uu}$}
\put(60,327){\small $0^{ud}$}
\put(175,348){$+$}
\put(200,345){\line(1,0){18}}
\put(200,347){\line(1,0){17.5}}
\put(200,349){\line(1,0){17}}
\put(200,351){\line(1,0){17}}
\put(200,353){\line(1,0){17.5}}
\put(200,355){\line(1,0){18}}
\put(230,350){\circle{25}}
\put(219,346){\line(1,1){15}}
\put(222,341){\line(1,1){17}}
\put(227.5,338.5){\line(1,1){14}}
\put(247.5,360){\circle{16}}
\put(247.5,340){\circle{16}}
\put(252,367){\vector(1,1){18}}
\put(252,333){\vector(1,-1){18}}
\put(252,367){\vector(1,-1){17}}
\put(252,333){\vector(1,1){17}}
\put(277,350){\circle{16}}
\put(285,350){\vector(1,1){18}}
\put(285,350){\vector(1,-1){18}}
\put(230,330){\circle{16}}
\put(230,322){\vector(1,-2){9}}
\put(230,322){\vector(-1,-2){9}}
\put(261,362){1}
\put(261,331){2}
\put(306,365){1}
\put(306,327){2}
\put(274,385){5}
\put(274,310){6}
\put(242,305){3}
\put(212,305){4}
\put(240.5,357){\small $1^{uu}$}
\put(240.5,337){\small $0^{db}$}
\put(270,347){\small $0^{ud}$}
\put(223,327){\small $1^{uu}$}
\put(327,348){$+$}
\put(345,348){2}
\put(363,345){\line(1,0){18}}
\put(363,347){\line(1,0){17.5}}
\put(363,349){\line(1,0){17}}
\put(363,351){\line(1,0){17}}
\put(363,353){\line(1,0){17.5}}
\put(363,355){\line(1,0){18}}
\put(393,350){\circle{25}}
\put(382,346){\line(1,1){15}}
\put(385,341){\line(1,1){17}}
\put(390.5,338.5){\line(1,1){14}}
\put(402,368){\circle{16}}
\put(402,332){\circle{16}}
\put(405,353){\vector(3,1){28}}
\put(405,347){\vector(3,-1){28}}
\put(410,370){\vector(3,2){21}}
\put(410,370){\vector(3,-1){23}}
\put(410,330){\vector(3,1){23}}
\put(410,330){\vector(3,-2){21}}
\put(441,360){\circle{16}}
\put(441,340){\circle{16}}
\put(450,360){\vector(3,2){21}}
\put(450,360){\vector(3,-1){23}}
\put(450,340){\vector(3,1){23}}
\put(450,340){\vector(3,-2){21}}
\put(477,371){1}
\put(477,352){2}
\put(478,342){3}
\put(477,319){4}
\put(421,386){5}
\put(421,307){6}
\put(426,368){1}
\put(426,351){2}
\put(426,342){3}
\put(426,325){4}
\put(395,365){\small $1^{uu}$}
\put(395,329){\small $0^{ub}$}
\put(434,357){\small $1^{uu}$}
\put(434,337){\small $0^{ud}$}
\put(10,248){$+$}
\put(24,248){2}
\put(37,245){\line(1,0){18}}
\put(37,247){\line(1,0){17.5}}
\put(37,249){\line(1,0){17}}
\put(37,251){\line(1,0){17}}
\put(37,253){\line(1,0){17.5}}
\put(37,255){\line(1,0){18}}
\put(67,250){\circle{25}}
\put(56,246){\line(1,1){15}}
\put(59,241){\line(1,1){17}}
\put(65.5,238.5){\line(1,1){14}}
\put(76,268){\circle{16}}
\put(76,232){\circle{16}}
\put(79,253){\vector(3,1){28}}
\put(79,247){\vector(3,-1){28}}
\put(84,270){\vector(3,2){21}}
\put(84,270){\vector(3,-1){23}}
\put(84,230){\vector(3,1){23}}
\put(84,230){\vector(3,-2){21}}
\put(115,260){\circle{16}}
\put(115,240){\circle{16}}
\put(124,260){\vector(3,2){21}}
\put(124,260){\vector(3,-1){23}}
\put(124,240){\vector(3,1){23}}
\put(124,240){\vector(3,-2){21}}
\put(151,271){1}
\put(152,252){2}
\put(152,242){3}
\put(151,219){4}
\put(95,286){5}
\put(95,207){6}
\put(100,268){1}
\put(100,251){2}
\put(100,242){3}
\put(100,225){4}
\put(69,265){\small $1^{uu}$}
\put(69,229){\small $0^{db}$}
\put(108,257){\small $1^{uu}$}
\put(108,237){\small $0^{ud}$}
\put(175,248){$+$}
\put(190,248){2}
\put(205,245){\line(1,0){18}}
\put(205,247){\line(1,0){17.5}}
\put(205,249){\line(1,0){17}}
\put(205,251){\line(1,0){17}}
\put(205,253){\line(1,0){17.5}}
\put(205,255){\line(1,0){18}}
\put(235,250){\circle{25}}
\put(224,246){\line(1,1){15}}
\put(227,241){\line(1,1){17}}
\put(232.5,238.5){\line(1,1){14}}
\put(254,257){\circle{16}}
\put(251,237){\circle{16}}
\put(261,261){\vector(1,1){15}}
\put(236,262.5){\vector(3,1){40}}
\put(284,277){\circle{16}}
\put(261,261){\vector(1,-1){12}}
\put(259,235){\vector(1,1){14}}
\put(281,250){\circle{16}}
\put(289,250){\vector(3,2){18}}
\put(289,250){\vector(3,-2){18}}
\put(292,278){\vector(3,2){18}}
\put(292,278){\vector(3,-2){18}}
\put(259,235){\vector(1,-1){16}}
\put(236,238){\vector(1,-2){11}}
\put(315,287){1}
\put(315,264){2}
\put(311,254){3}
\put(311,232){4}
\put(265,280){1}
\put(270,260){2}
\put(262,246){3}
\put(265,232){4}
\put(278,211){5}
\put(250,209){6}
\put(247,254){\small $0^{ud}$}
\put(244,234){\small $1^{uu}$}
\put(277,274){\small $0^{ud}$}
\put(274,247){\small $1^{uu}$}
\put(337,248){$+$}
\put(355,248){2}
\put(370,245){\line(1,0){18}}
\put(370,247){\line(1,0){17.5}}
\put(370,249){\line(1,0){17}}
\put(370,251){\line(1,0){17}}
\put(370,253){\line(1,0){17.5}}
\put(370,255){\line(1,0){18}}
\put(400,250){\circle{25}}
\put(389,246){\line(1,1){15}}
\put(392,241){\line(1,1){17}}
\put(397.5,238.5){\line(1,1){14}}
\put(419,257){\circle{16}}
\put(416,237){\circle{16}}
\put(426,261){\vector(1,1){15}}
\put(401,262.5){\vector(3,1){40}}
\put(449,277){\circle{16}}
\put(426,261){\vector(1,-1){12}}
\put(424,235){\vector(1,1){14}}
\put(446,250){\circle{16}}
\put(454,250){\vector(3,2){18}}
\put(454,250){\vector(3,-2){18}}
\put(457,278){\vector(3,2){18}}
\put(457,278){\vector(3,-2){18}}
\put(424,235){\vector(1,-1){16}}
\put(401,238){\vector(1,-2){11}}
\put(480,287){1}
\put(480,264){2}
\put(476,254){3}
\put(476,232){4}
\put(430,280){1}
\put(435,260){2}
\put(427,246){3}
\put(430,232){4}
\put(443,211){5}
\put(415,209){6}
\put(412,254){\small $1^{uu}$}
\put(409,234){\small $0^{db}$}
\put(442,274){\small $1^{uu}$}
\put(439,247){\small $0^{ud}$}
\put(10,148){$+$}
\put(28,148){2}
\put(45,145){\line(1,0){18}}
\put(45,147){\line(1,0){17.5}}
\put(45,149){\line(1,0){17}}
\put(45,151){\line(1,0){17}}
\put(45,153){\line(1,0){17.5}}
\put(45,155){\line(1,0){18}}
\put(75,150){\circle{25}}
\put(64,146){\line(1,1){15}}
\put(67,141){\line(1,1){17}}
\put(72.5,138.5){\line(1,1){14}}
\put(84,168){\circle{16}}
\put(84,132){\circle{16}}
\put(95,150){\circle{16}}
\put(103,150){\vector(1,1){12}}
\put(103,150){\vector(1,-1){12}}
\put(92,170){\vector(3,2){21}}
\put(92,170){\vector(3,-1){23}}
\put(92,130){\vector(3,1){23}}
\put(92,130){\vector(3,-2){21}}
\put(123,160){\circle{16}}
\put(123,140){\circle{16}}
\put(132,160){\vector(3,2){21}}
\put(132,160){\vector(3,-1){23}}
\put(132,140){\vector(3,1){23}}
\put(132,140){\vector(3,-2){21}}
\put(159,171){1}
\put(160,152){2}
\put(160,142){3}
\put(159,119){4}
\put(103,186){5}
\put(103,107){6}
\put(108,168){1}
\put(102,156){2}
\put(102,137){3}
\put(108,125){4}
\put(77,165){\small $1^{uu}$}
\put(77,129){\small $0^{db}$}
\put(88,147){\small $1^{uu}$}
\put(116,157){\small $1^{uu}$}
\put(116,137){\small $0^{ud}$}
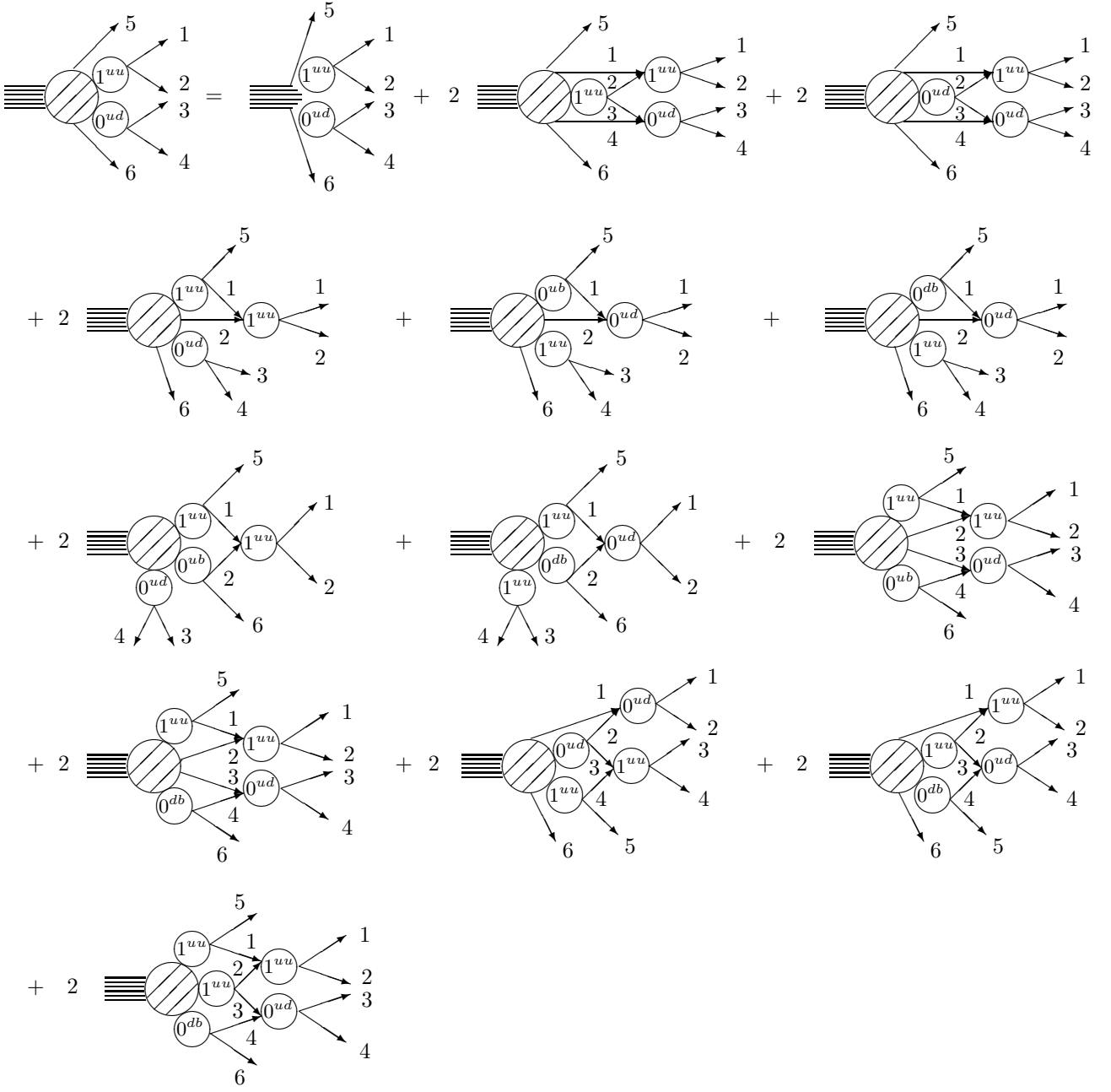
\put(0,70){Fig. 1. Graphical representation of the amplitude $A_2^{1^{uu}0^{ud}}$ for the
case of $\Delta\Lambda_b$, $I=\frac{3}{2}$, $J^P=2^+$.}
\end{picture}

\end{document}